# HIV and TB in Eastern and Southern Africa:
# Evidence for behaviour change and the impact of ART

Brian G. Williams

South African Centre for Epidemiological Modelling and Analysis (SACEMA), Stellenbosch, South Africa
Correspondence to BrianGerardWilliams@gmail.com

**Abstract**

The United Nations Joint Programme on HIV/AIDS (UNAIDS) has set a target to ensure that 15 million HIV-positive people in the world would be receiving combination anti-retroviral treatment (ART) by 2015. This target is likely to be reached and new targets for 2020 and 2030 are needed. Eastern and Southern Africa (ESA) account for approximately half of all people living with HIV in the world and it will be especially important to set reachable and affordable targets for this region. In order to make future projections of HIV and TB prevalence, incidence and mortality assuming different levels of ART scale-up and coverage, it is first necessary to assess the current state of the epidemic. Here we review national data on the prevalence of HIV, the coverage of ART and the notification rates of TB to provide a firm basis for making future projections. We use the data to assess the extent to which behaviour change and ART have reduced the number of people living with HIV who remain infectious.

## Introduction

In order to assess the current status and trends in the prevalence of HIV, the coverage of ART and the notification rates of TB for all the countries in the UNAIDS Eastern and Southern Africa (ESA) region we start from the UNAIDS model fits[1] for HIV and ART and the WHO data for TB.[2] We fit the HIV data to a double-logistic function and use the fitted function to estimate the impact on the TB epidemic assuming a variable delay and incidence rate ratio for TB in people with HIV, with or without ART. Where the data appear to be inconsistent we refer back to the original data on which the UNAIDS model fits were made and adjust the estimates accordingly. This gives our best estimates of the current status and trends for HIV and TB. It is important to stress that this is not a transmission model so that it does not allow for reductions in transmission arising from the provision of ART. The purpose of this study is to ensure that we have the best estimates of the current trend data which can then be used to drive a dynamical transmission model.

## Analysis

For the HIV epidemics we fit a double-logistic function

$$P(t) = a_1 \frac{e^{r_1(t-t_1)}}{1-e^{r_1(t-t_1)}} - a_2\left(1-\frac{e^{r_2(t-t_2)}}{1-e^{r_2(t-t_2)}}\right) \quad 1$$

to the reported prevalence data. The first term in Equation 1 accounts for the rate, $r_1$, the timing $t_1$, and the peak prevalence $a_1$ of the initial rise in the prevalence. The second term accounts for the rate, $r_2$, the timing $t_2$, and the subsequent drop, $a_2$, in the prevalence as a result of reductions in transmission not associated with the increasing coverage of ART. While we refer to this as 'behaviour change' the reasons for and drivers of these reductions in transmission are not understood and we discuss this further below.

For the ART coverage we fit a single logistic function

$$A(t) = a_3 \frac{e^{r_3(t-t_3)}}{1-e^{r_3(t-t_3)}} \quad 2$$

to the reported coverage data where the rate, $r_3$, the timing $t_3$, and the asymptotic prevalence of ART $t_3$, account for the increasing coverage of ART.

For the TB notification rates we use three parameters: the pre-HIV notification rate $n$, the time delay between the rise in ART prevalence and TB notifications, $\tau$, and the incidence rate ratio for TB when people are infected with HIV, $\eta$. We assume that the incidence rate ratio for TB when people are on ART is $0.4\eta$.[3] The TB notification rate is then

$$T(t) = n(1-P(\tau-\tau)) + \\ n\{\eta(P(t-\tau)-A(t-\tau))+0.4\eta A(t-\tau)\} \quad 3$$

## Results

Figure 1 gives the data for HIV and TB for each of the countries in ESA and the fitted functions (Equations 1–3). For HIV the red dots are the UNAIDS estimates of the prevalence of HIV; the red lines are the fitted values. The green dots are the UNAIDS estimates of the coverage of ART; the green lines are the fitted values. The blue lines are the red lines minus the green lines and give estimates of the prevalence of HIV for people who are not on ART. For HIV the denominator is the total adult population aged fifteen years or more.

For TB the green lines give the notification rate in HIV-negative people, the blue lines include the notification rate for people infected with HIV but not on ART, the red lines include the notification rate for people infected with HIV and on ART, that is to say the total notification rate. For TB the denominator is the total population.

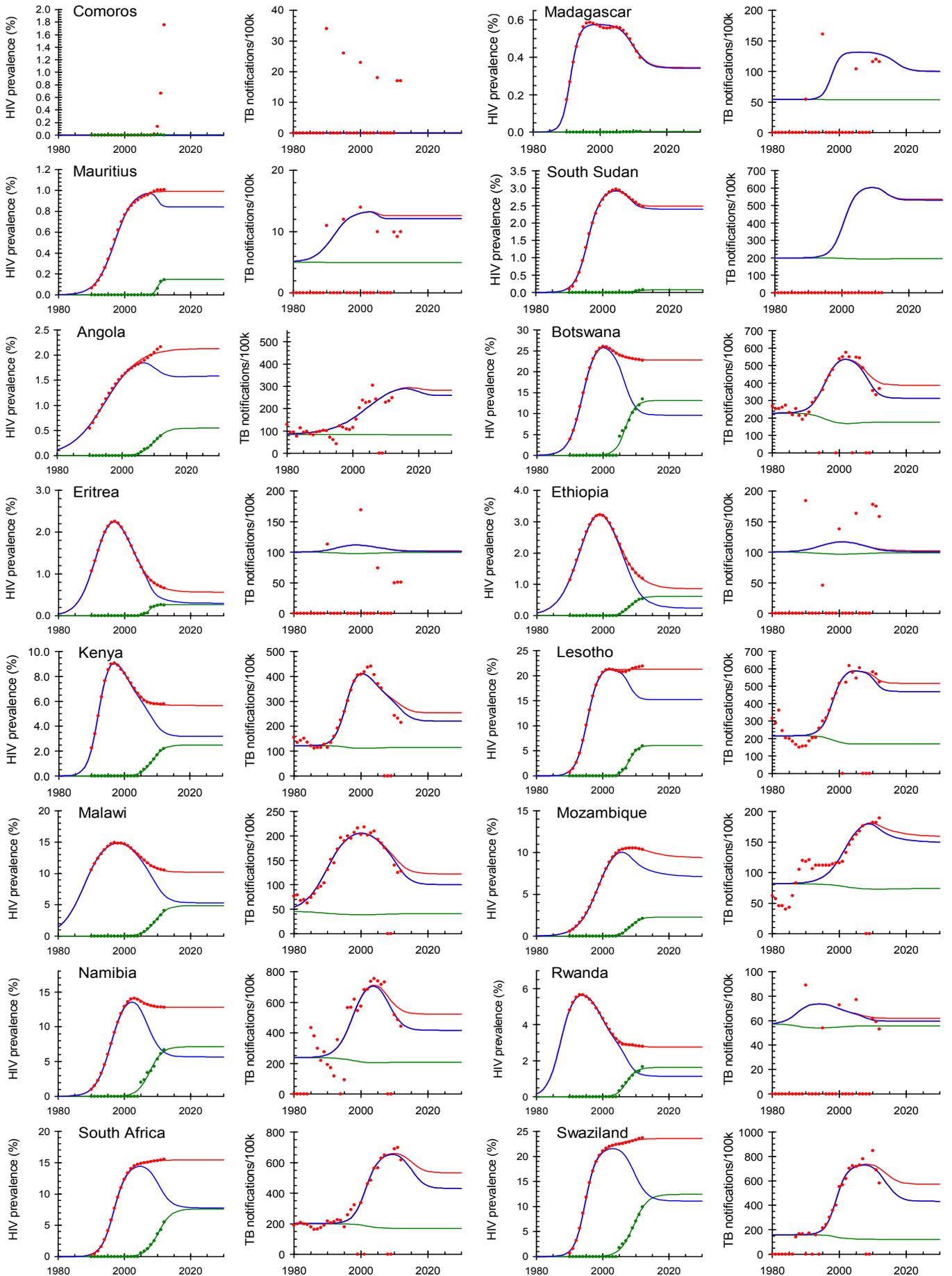

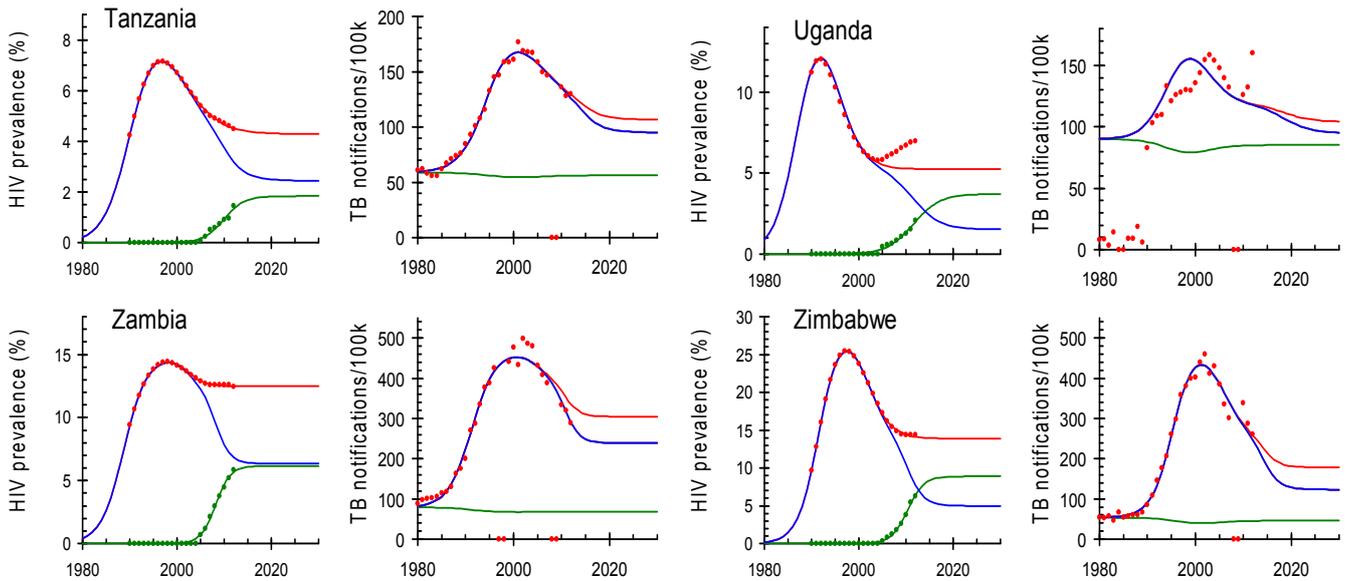

Figure 1. HIV prevalence among adults 15+ years old and TB notifications per 100k population for the countries in ESA. HIV: UNAIDS estimates of HIV prevalence (red dots) and ART prevalence (green dots). Fitted total prevalence, red lines, prevalence of ART (green lines) prevalence for those not on ART (blue lines). TB: WHO estimates of TB notification rates (red dots). Fitted notification rates for HIV negative people (green lines); including HIV positive people not on ART (blue lines); including HIV people on ART, that is total notification rate, (red lines).

The data for the Comoros in Figure 1 are insufficient to estimate the trends in either HIV or TB. The HIV data for Madagascar show a flat peak followed by a decline which is unlike the trend in any other country and this needs to be examined further. The data for Angola show a steady rise unlike that in any other country and are considered in more detail below. The other countries show either a rise to a more or less constant asymptote or a peak followed by a decline that precedes the roll-out of ART and which we attribute to a reduction in the risk of infection due to an unspecified change in behaviour. In South Africa, Swaziland and Lesotho the prevalence flattens off but continues to rise after the year 2000. This slow rise is surprising in view of the ante-natal clinic data for South Africa which show a flattening of the prevalence of HIV in all nine provinces and no indication of a continuing increase.[4] Here we assume that the prevalence reaches a plateau but this too needs further consideration. The data for Uganda show a rapid decline followed by a significant upturn in the prevalence and this recent increase needs to be examined carefully. If it were the case that the decline in the prevalence of HIV in Uganda came about because an increased awareness of AIDS related mortality then the subsequent fall in mortality could produce the upturn as shown in a modelling study for Zimbabwe.[5]

We first examine the raw data for Angola. The data on which the UNAIDS trend estimates are based (Appendix 1; Table 1) show the sparse nature of the data for Angola. In particular there is one early data point for Zaire Province where a prevalence of 9% in was recorded in 1988 but no further data were available until 2007 after which time the prevalence remains below 2.3%. It seems prudent to exclude this point.

This is followed by a series of estimates ranging from 5% to 8% in Cabinda between 1992 and 1999. However, Cabinda is an enclave separated from Angola by the Democratic Republic of the Congo so it seems safe to exclude these data also. The only data that allow us to estimate the timing and rate of the initial rise in HIV are from four data points in Luanda between 1995 and 2001 and if we use these to set the timing and the rate of the initial increase we get the fits shown in Figure 2 which appear to be more reasonable than the fits for Angola given in Figure 1.

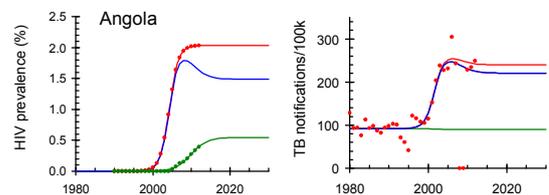

Figure 2. HIV prevalence (%) and TB notifications (per 100k population) for Angola. As described in the text this excludes the data for Cabinda and Zambia and uses the data from Luanda to determine the rate and timing of the initial rise in the prevalence of HIV.

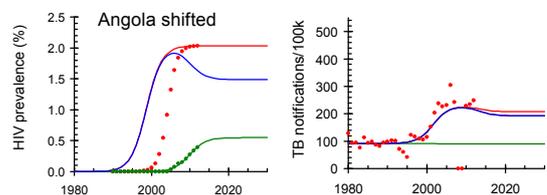

Figure 3. The fitted HIV prevalence (% of adults 15+ years old) and TB notifications (per 100k population) for Angola. For the reasons discussed in the text we use the fit to the TB data to determine the rise in HIV prevalence assuming a delay between the two of 3 years.

This still raises questions about the time trend of the prevalence of HIV in Angola. We note first (Figure 2 c.f. Figure 1) that the epidemic of HIV in Angola appears to have risen about five years later than in any other country; that the initial rate of increase is 30% greater than in any other country and the rise in TB predates the rise in HIV by 2.7 years which cannot be the case.

Since the timing was based on only four points for Luanda we therefore set the initial rate of increase to 0.5/year (the average of all the other countries in the region) and set the delay between the rise of HIV and the rise of TB to 3 years. This gives the fits shown in Figure 3 and we use these to describe the trends in HIV and TB for Angola. We now consider the values of the fitted parameters used in Figure 1, using Figure 3 for Angola, and these are given in Figure 4.

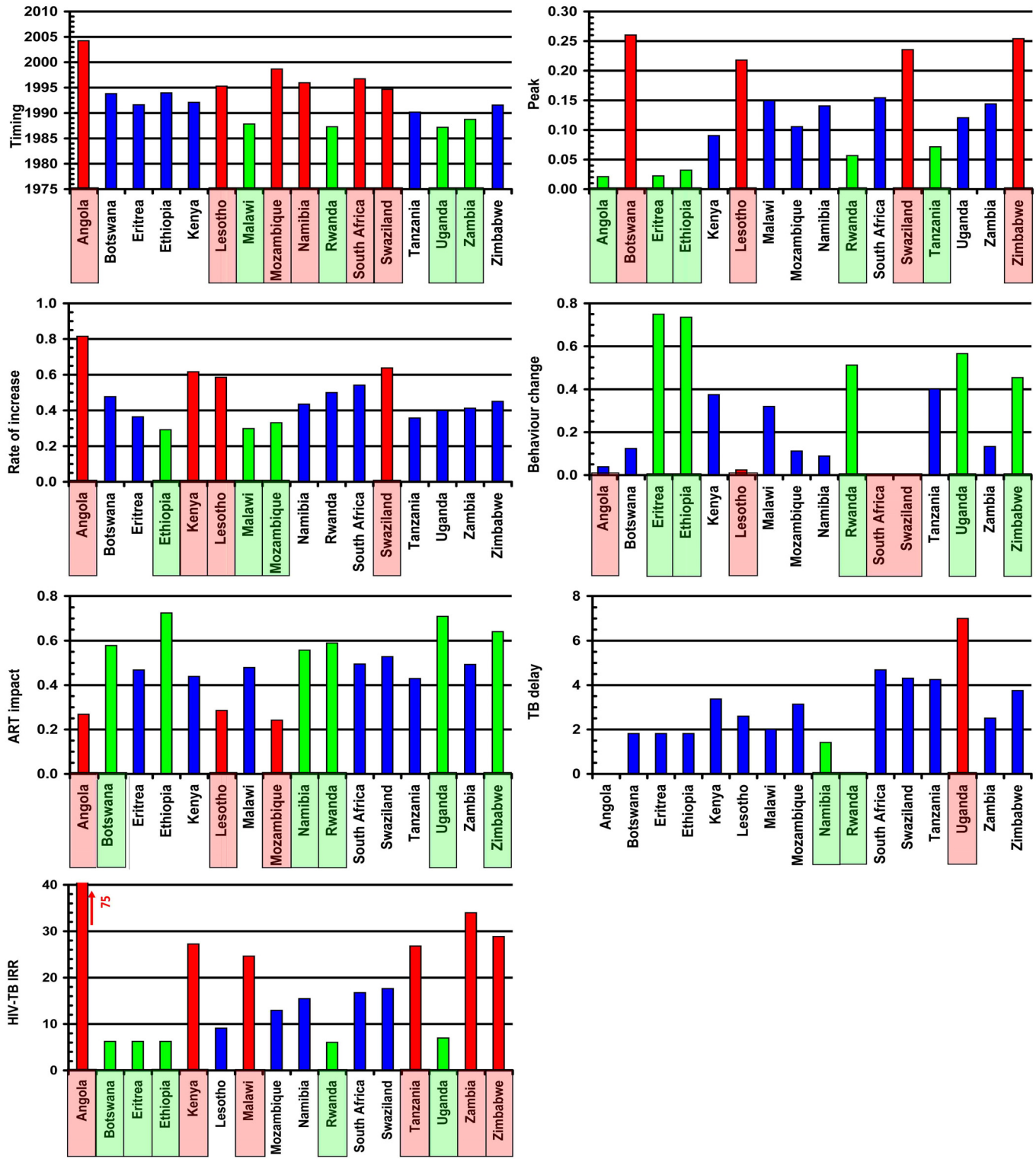

Figure 4. Parameters for the fits to the HIV and TB data for the countries in ESA. Top left: The timing of the epidemic as measured by the year in which the initial rise reaches half of its asymptotic value. Top right: the peak prevalence of HIV. Row 2, left: The initial rate of increase/year. Row 2, right: The proportional reduction in the prevalence of HIV resulting from reductions in transmission preceding the roll-out of ART. Row 3, left: The proportional reduction in the prevalence as a result of the roll-out of ART. Row 3, right: The time delay between the rise of HIV and the rise of TB. Bottom: The estimated values of the incidence rate ratio.

**Discussion**

The fits in Figure 1, Figure 3 for Angola, provide the current best estimates of the trend in HIV, ART and TB for the countries of the ESA region and should be used to inform dynamical models of HIV and TB with a view to projecting future trends and assessing the impact and costs of different combinations of interventions. This preliminary data cleaning process reveals important insights into the nature of the epidemics in ESA which can also be used to guide further analysis.

With regard to timing (Figure 4, top left) the epidemic of HIV started first in the East African countries of Uganda, Rwanda, Zambia and Malawi and last in Lesotho, Mozambique, Namibia, Swaziland and South Africa. This is consistent with the notion that HIV arose first in Central Africa and then spread from there to East and then to southern Africa. It is likely that the restriction of movement and the isolation of South Africa under the Apartheid Regime delayed the spread of HIV into South Africa and some of the neighbouring states and the war in Mozambique, which ended in 1992[1] may have further delayed the spread of HIV in that country.

The peak prevalence of HIV in the ESA Region (Figure 4, top right) varies by an order of magnitude. Excluding the island countries the lowest prevalence rates of HIV were observed in Angola, Eritrea and Ethiopia; the highest in Botswana, Lesotho, Swaziland and Zimbabwe. While Swaziland is closely connected to KwaZulu/Natal, where the peak prevalence was even higher, Lesotho is heavily dependent on remittances from mine workers in South Africa. The reasons for the high rates of HIV in Botswana and Zimbabwe are less clear. While Batswana are employed in significant numbers to work on the mines in South Africa, men from Zimbabwe were only ever employed on the gold mines in South Africa in small numbers.

It should be noted that the UNAIDS data suggest that in South Africa, Lesotho and Swaziland, the prevalence of HIV is still increasing slowly. However, the prevalence of HIV among ante-natal clinic women in South Africa has been flat for the last five years in all nine provinces so we will assume that the apparent slow rise is an artefact of the UNAIDS model in South Africa, Lesotho and Swaziland.

The countries in which the initial rate of spread was fastest (Figure 4, second row, left) were Lesotho, Swaziland, South Africa and Kenya. We exclude Angola as the early data there are unreliable. The rapid rise in Lesotho and Swaziland may have been driven by the considerable extent of migration between these two countries and South Africa. The reasons for the rapid rise in Kenya are less clear but it should be noted that about 30% of all cases of HIV occur in nine districts in South Nyanza that border on Lake Victoria[6] and in these nine districts about 20% of women attending ante-natal clinics were infected with HIV in 2006, more than twice the rate in any other district including Mombassa and Nairobi. The countries in which the initial rate of increase was slowest were Eritrea, Ethiopia, Malawi, Mozambique and Tanzania.

In many, but not all, countries the prevalence of infection fell significantly before the roll-out of ART began. The greatest proportional declines (Figure 4, second row, right) were seen in Eritrea and Ethiopia and Rwanda but the peak prevalence in these countries was low. Uganda and Zimbabwe are of particular interest; the peak prevalence was high, 12% and 25%, respectively, but then fell rapidly by about 50% before ART became available for reasons which are still not clear. In South Africa, Lesotho and Swaziland there is no evidence of any decline in prevalence prior to the roll-out of ART.

The scale up of ART in ESA has been impressive (Figure 4, third row left) leading to a further decline in the prevalence of HIV among people not on ART ranging from of about 25% in Angola, Lesotho and Mozambique to 65% in Ethiopia, Uganda and Zimbabwe. In Zimbabwe the combined effect of behaviour change and the availability of ART led to an 80% drop in the prevalence of people not on ART; almost enough to ensure elimination.

The data on TB are also informative. There is clearly a problem with Uganda where the rise in the TB notification rate (Figure 4, third row right) followed the rise in HIV prevalence by almost seven years. However the TB data for Uganda are weak and variable. In the other countries in the region the rise in TB follows the rise in HIV by about 2 to 4 years although it is less than 2 in Namibia and the data for Rwanda are too uncertain to draw firm conclusions.

Finally, we note that the incidence rate ratio for TB varies greatly among countries (Figure 4, bottom) for reasons that are not understood. In Botswana, Eritrea, Ethiopia and Rwanda the incidence rate ratio is about 5; in Kenya, Malawi, Tanzania, Zambia and Zimbabwe it ranges from 25 to 35.

**Conclusions**

The data on HIV and TB in ESA are quite robust for most countries. Some countries need to review their data and all countries could expand their monitoring and routine surveillance systems to great advantage.

Probably the best surveillance in ESA is the series of annual surveys, starting in 1990, for women attending ante-natal clinics in South Africa.[4] However, if these samples were tested, retrospectively for the presence of anti-retroviral drugs, viral loads, drug resistance and recent infections, it would be possible to assess the current state of the epidemic with much greater confidence and to make projections with much greater certainty.

All countries need to strengthen their patient monitoring systems and in this regard the Malawi Quarterly reports provide a good example and instructive lessons on how it should be done.[7] For the island countries, where the prevalence is very low, better data are needed on the prevalence and size of the 'key populations' which are probably driving the epidemic in those places.

This analysis depends on the validity of the estimates of ART coverage in each country. However, it will be increasingly important to distinguish between the cumulative number of people of have ever started treatment and the number that remain on treatment at any given time. Very few data are available anywhere in the region to give us great confidence in these numbers and suitable surveillance should be put in place to remedy this situation.

---

1  http://en.wikipedia.org/wiki/Mozambican_Civil_War.

**Comoros, Seychelles, South Sudan**

Very limited data are available for either HIV or TB in the Comoros (Figure 1), the Seychelles or South Sudan. Better surveillance data for both infections and for key populations are needed.

**Madagascar**

There are better data on HIV in Madagascar (Figure 1) but the UNAIDS model suggests the prevalence was very low, always below about 0.6%, and flat from 1995 to 2007, after which it began to decline rapidly. This is quite unlike the pattern seen in any other country in ESA and deserves further investigation. About 1% of people infected with HIV are on ART. The TB data are limited and better surveillance for TB is needed. Better surveillance data for both infections and for key populations are needed.

**Mauritius**

The adult prevalence of HIV in Mauritius (Figure 1) has remained below about 1% but shows no sign of decline. About 14% of people infected with HIV are on ART and if this is expanded it should be possible to keep the prevalence at very low levels. The TB data are limited and better surveillance for TB is needed. Better surveillance data for both infections and for key populations are needed.

**Angola**

The revised data for HIV in Angola (Figure 3) suggest that the epidemic started later than in most other countries in ESA, has remained below about 2.5% but shows no sign that there has been any significant change in risky behaviour. About 19% of people infected with HIV are on ART and coverage needs to be expanded. Given the low prevalence of HIV the rise in TB notification rates seems to be disproportionately high. This should be investigated further but if relatively small numbers of people in concentrated areas are a risk of both infections, this would artificially inflate the incidence rate ratio for TB in those with HIV.[8]

**Botswana**

Botswana has experienced the most severe epidemic in ESA with a peak prevalence of HIV in the general population of about 25%. There seems to be little evidence of changes in behaviour that would have lead to a reduction in HIV transmission but the roll-out of ART has been very successful and about 60% of people infected with HIV are on ART making it the highest in the region. This will have substantially reduced the incidence of AIDS related TB by about 50%. Botswana now needs to move towards earlier treatment for HIV.

**Eritrea**

The epidemic of HIV in Eritrea peaked at just over 2% and the data suggest that there was a substantial change in behaviour which brought down the prevalence by about 75% and about 40% of people infected with HIV are on ART. However, the data underlying the UNAIDS estimates are weak and better data surveillance is needed. Finding the relatively few remaining cases will be challenging and the immediate need is to find out who the at-risk people are and develop efficient strategies to find them and offer them treatment. It should be possible to end AIDS in Eritrea.

**Ethiopia**

The situation in Ethiopia is similar to that in Eritrea. The peak prevalence was about 3%, changes in behaviour seem to have reduced transmission by about 75% the about 48% of people infected with HIV are on ART. As in Eritrea the challenge is to find the relatively few remaining cases and develop efficient strategies to find them and offer them treatment. It should also be possible to end AIDS in Ethiopia.

**Kenya**

The epidemic of HIV in Kenya spread rapidly in the early years and peaked at an adult prevalence of about 9%. Behaviour change in Kenya seems to have reduced the prevalence by about 37% and about 38% of people infected with HIV are on ART. This has, in turn, reduced AIDS related TB by about 50%. It is of particular interest to note that nine districts in Kenya, bordering on Lake Victoria but excluding Busia, account for 10% of the Kenyan population but 30% of the HIV infections[6] and successful control of HIV will depend on how HIV is dealt with in these nine districts.

**Lesotho**

Lesotho is a small country heavily dependant on South Africa socially and economically. The peak prevalence of HIV was 22%, the epidemic started late but then spread very rapidly. There is no evidence that changing behaviour has had any impact on HIV but 28% of people infected with HIV are now on ART and there are indications that AIDS related TB is beginning to fall. A key issue for Lesotho is migration to and from South Africa, especially in relation to the gold and platinum mines, and successful control of HIV will depend on control in South Africa.

**Malawi**

Malawi was among the first countries in ESA to experience the epidemic of HIV. It is of interest to note that in 1987 HIV testing was done on workers on the South African gold mines in which correlations were made between their HIV status and their country of origin.[9] The men in the study came from South Africa, Malawi, Lesotho, Mozambique, Swaziland and Botswana but notably not from Rhodesia, now Zimbabwe. The prevalence amongst mine-workers from Malawi was 4%; among those from all other countries, including South Africa, it was about 0.03%. As a result the Chamber of Mines stopped recruiting novices from Malawi[10] and the number of Malawians employed on the South African mines fell from 13,090 in 1988 to 2,212 in 1989.[11] The peak prevalence of HIV in Malawi was about 15% but changes in behaviour reduced the prevalence of HIV by about 32% and about 38% of people infected with HIV are on ART. Malawi has a very good patient monitoring system[7] but needs to expand access to ART.

**Mozambique**

The epidemic of HIV in Mozambique started later than in other countries in ESA and reached a peak prevalence of 11%. There seems to have been a small reduction in the prevalence as a result of changes in behaviour and 20% of all people infected with HIV are on ART. While this is a good start, ART needs to be made much more widely available. Furthermore, the recent and currents efforts to open up large areas of northern Mozambique to mining could increase the spread of HIV and this should be considered and monitored very carefully.

**Namibia**

The epidemic of HIV in Namibia started late and reached a peak prevalence of 14%. Behaviour change appears to have reduced the prevalence of HIV by about 9%. There has been a successful roll-out of ART and about 51% of people infected with HIV are on ART. With a relatively small population concentrated in relatively small geographical regions of the country a substantial increase in the roll-out of ART should bring the epidemic in Namibia under control.

**Rwanda**

The epidemic of HIV in Rwanda started early but was relatively modest a with peak prevalence just below 6%. Behaviour change appears to have reduced the prevalence by about 51% and the roll-out of ART has been very successful with 56% of infected people being on ART. Unfortunately the TB data are unreliable. However, if ART coverage is expanded Rwanda should be in a good position to end AIDS.

**South Africa**

South Africa accounts for about 16% of all people living with HIV in the world and control in the world depends critically on control in South Africa. The epidemic reached a peak prevalence of 15% but ranging from about 8% in the Western Cape to about 26% in KwaZulu/Natal. The epidemic started later than in the rest of the world but once HIV was established the rate of spread was relatively fast. There is no evidence of any reduction in HIV prevalence arsing from changes in behaviour but about 35% of people living with HIV are now on ART. While South Africa has the biggest ART programme in the world, further steps are needed to increase ART coverage further. There are indications that ART is already beginning to reduce the TB notification rate which has fallen by about 25%. Given the lack of evidence that changes in behaviour have had any impact on HIV in South Africa an ambitious policy of early treatment will be needed to control the epidemic.

**Swaziland**

The epidemic of HIV in Swaziland is dependent on the epidemic in South Africa but has reached an even higher peak prevalence of HIV of 25% and HIV spread even more quickly than in South Africa. As in South Africa there is no evidence for any reduction in the prevalence of HIV arising from changes in behaviour but the provision of ART has had a substantial impact and 42% of all people living with HIV are on ART. There is some evidence that ART is beginning to reduce the TB notification rate. As in South Africa an ambitious policy of early treatment will be needed to control the epidemic.

**Tanzania**

The epidemic of HIV in Tanzania peaked at about 7% but there is evidence that changes in behaviour reduced HIV prevalence by about 40% and 29% of people living with HIV are not on ART. With further expansion of ART Tanzania should be able to control the epidemic.

**Uganda**

Uganda was one of the first countries in the world to experience the full force of the HIV epidemic and the prevalence rose to a peak of about 12%. However, there was early evidence of a substantial change in people's behaviour leading to a decline in the prevalence of HIV of about 57% and about 38% of people living with HIV are on ART. However, Uganda remains the only country in the region with evidence of a significant upsurge in the prevalence of HIV and it is important that this is confirmed and, if found to be a real, understood. At present the best explanation is that as people became aware of AIDS related mortality they were less likely to engage in high-risk sex but as the prevalence fell, and with it the AIDS related deaths, people became more complacent about high-risk sex. Modelling studies[5] show that this could produce oscillations of precisely this kind but the question demands further investigation.

**Zambia**

In Zambia the prevalence of HIV peaked at about 14% and there seems to have been some change in behaviour leading to a 13% drop in the prevalence of HIV. The scale up of HIV has been relatively successful and 45% of people living with HIV are now on ART. Zambia has now adopted a policy of early treatment and if this is effective there is every reason to believe that Zambia will be able to control the epidemic of HIV.

**Zimbabwe**

The epidemic of HIV in Zimbabwe reached the second highest levels in the region, just after Botswana, with a prevalence of HIV of 25% in 1997. However, there appears to have been a very rapid and substantial response with behaviour change leading to a 45% reduction in prevalence and 45% of people living with HIV are not on ART. This alone should bring the epidemic close to the elimination threshold and if the availability of ART is increased further, elimination of HIV in Zimbabwe should be assured. Because the response to the epidemic of HIV in Zimbabwe happened so early there is clear evidence that this has led to a 50% decline in AIDS related TB.

## Desiderata

Progress in responding to and managing HIV in ESA has been considerable even though the region contains some of the poorest countries with the highest rates of HIV in the world (Figure 4, top left). The roll-out of ART has been particularly striking and this has led to a reduction in the prevalence of people living with HIV who are not on ART, and consequently of the incidence of HIV, of about 50% but ranging from about 25% to 75% (Figure 4, third row, left). But many puzzles remain and much still needs to be done.

In some countries there is clear evidence that the prevalence of HIV began to fall rapidly well before ART became available or prevention programmes, such as medical male circumcision or vaginal microbicides, were in place.[5] Epidemiological models show that this decline in the number of people living with HIV was too great to be the result of people dying of AIDS. There must have been changes in people's behaviour or in the social structure and networks that facilitate the spread of infection to account for this observation. It is equally clear that in other countries such changes did not occur. At the moment there is no convincing explanation as to what these changes were, what factors facilitated or hindered them, and why the response was so variable across the region. For want of a better understanding we have gathered these effects into the overall rubric of 'behaviour change'.

We need to understand the geographical heterogeneity in infections rates. This is illustrated most directly in Kenya where a 2006 survey of about half-a-million women attending ante-natal clinics showed that there are nine districts, bordering Lake Victoria, which have 10% of the total population but 30% of the people living with HIV.[6]

A further, equally puzzling aspect of the epidemic is the variability in the incidence rate ratio for TB if people have HIV. Being infected with HIV increases the risk of getting TB by about 30 times in Zambia and Zimbabwe but only by about 5 times in Botswana, Rwanda and Uganda. This dramatic difference is in urgent need of explanation.

What then still needs to be done? First, much better surveillance data on the prevalence of HIV *and* ART, of the incidence of HIV, and of the extent of viral load suppression in those on ART is needed. Second, much better patient monitoring isx needed in order to understand and, if needs be, address problems associated with the ART treatment cascade and ensure acceptability of testing, and high levels of treatment uptake, compliance and viral load suppression. Third, these data should be used to estimate and better understand the current state of the epidemics, to make projections into the future, to explore the consequences of different intervention strategies for managing and ultimately controlling HIV. All of this must include a good understanding of all of the costs to the patients, to the health services, to the work-force and to the state. Fourth, convincing strategies, that be sold to politicians and funding agencies, must be developed so that the epidemic of AIDS can be ended.[12]

## Appendix 1

Table 1. The raw data for Angola that were used by UNAIDS to estimates the time-trend in the prevalence of HIV. The data give the percent prevalence of HIV in women attending ante-natal clinics in different years and provinces

| Clinic | Urb/Rur | Province | 1988 | 1989 | 1990 | 1991 | 1992 | 1993 | 1994 | 1995 | 1996 | 1997 | 1998 | 1999 | 2000 | 2001 | 2002 | 2003 | 2004 | 2005 | 2006 | 2007 | 2008 | 2009 | 2010 | 2011 | 2012 |
|---|---|---|---|---|---|---|---|---|---|---|---|---|---|---|---|---|---|---|---|---|---|---|---|---|---|---|---|
| Maternidade do Ca | urban | Bengo | | | | | | | | | | | | | | | | | 1.2 | 1.8 | | 4.5 | 3.3 | 3.0 | | | |
| Benguela | urban | Benguela | | | | | | | | | | | | | | | | | 0.6 | 2.8 | | 2.4 | 5.0 | 4.6 | | | |
| Maternidade do Lo | urban | Benguela | | | | | | | | | | | | | | | | | 0.6 | 2.6 | | 3.6 | 3.8 | 3.8 | | | |
| H.Prov.Kuito (Mate | urban | Bie | | | | | | | | | | | | | | | | | 0.6 | 0.8 | | 1.6 | 3.2 | 5.8 | | | |
| C.M.I. Cabinda | urban | Cabinda | | | | | | | | | | | | | | | | 3.0 | 3.2 | 2.8 | | 3.0 | 2.6 | 5.6 | | | |
| Cabinda Province | urban | Cabinda | | | | | | 6.0 | 7.0 | 7.0 | 5.0 | 8.0 | | | 8.0 | | | | | | | | | | | | |
| Cacongo | rural | Cabinda | | | | | | | | | | | | | | | | | | | | 3.4 | 0.9 | | 1.0 | 1.0 | |
| Cahama | rural | Cunene | | | | | | | | | | | | | | | | | | | | 2.4 | 2.6 | | 3.6 | 3.6 | |
| H.Prov.Ondjiva(Ma | urban | Cunene | | | | | | | | | | | | | | | 12.0 | | 8.8 | 10.6 | | 9.4 | 7.4 | 8.2 | | | |
| Ombanja | rural | Cunene | | | | | | | | | | | | | | | | | | | | 4.0 | 3.2 | 2.4 | | 2.4 | |
| Maternidade do Hu | urban | Huambo | | | | | | | | | | | | | | | | | 2.4 | 1.8 | | 3.6 | 4.2 | 3.4 | | | |
| Matala | rural | Huila | | | | | | | | | | | | | | | | | | | | 1.2 | 1.6 | 2.4 | | 2.4 | |
| Maternidade do Lu | urban | Huila | | | | | | | | | | | | | 4.0 | 1.0 | 2.6 | 4.0 | | 3.0 | 2.8 | 4.4 | | | | | |
| H.Prov.Menongue( | urban | Kuando Kubango | | | | | | | | | | | | | | | | | 4.0 | 3.0 | | 4.8 | 4.2 | 5.6 | | | |
| H.Central Ndalatan | urban | Kwanza Nord | | | | | | | | | | | | | | | | | 1.0 | 1.2 | | 1.6 | 1.0 | 2.0 | | | |
| Centro Policlínico | urban | Kwanza Sul | | | | | | | | | | | | | | | | | 0.8 | 1.4 | | 1.4 | 1.0 | 1.4 | | | |
| Gabela | rural | Kwanza Sul | | | | | | | | | | | | | | | | | | | | 2.6 | 1.0 | 0.4 | | 0.4 | |
| C. S. Viana (Ana Pa | urban | Luanda | | | | | | | | | | | | | | | | | 2.2 | 2.8 | | 1.8 | 3.0 | 4.8 | | | |
| C.S Ilha | urban | Luanda | | | | | | | | | | | | | | | | | | | | 2.9 | 3.4 | 5.2 | | | |
| C.S Samba | urban | Luanda | | | | | | | | | | | | | | | | | | | | 3.8 | 4.6 | 5.4 | | | |
| C.S. Asa Branca | urban | Luanda | | | | | | | | | | | | | | | 4.0 | | 2.0 | 1.8 | | 0.6 | 1.6 | 2.4 | | | |
| C.S. Cacuaco | urban | Luanda | | | | | | | | | | | | | | | | | 1.8 | 1.2 | | 1.8 | 2.2 | 2.6 | | | |
| C.S. Hoji ya Henda | urban | Luanda | | | | | | | | | | | | | | | 2.0 | | 3.8 | 4.4 | | 1.8 | 3.2 | 1.4 | | | |
| H. Kilamba Kiaxi | urban | Luanda | | | | | | | | | | | | | | | | 3.0 | 3.9 | 2.8 | | 4.0 | 3.0 | 2.6 | | | |
| Luanda | urban | Luanda | | | | | | | | 1.0 | | | | 3.0 | 3.0 | | 8.0 | | | | | | | | | | |
| Lucrecia Paim | urban | Luanda | | | | | | | | | | | | | | | | | 4.4 | 2.0 | | | | | | | |
| Maternidade Ngang | urban | Luanda | | | | | | | | | | | | | | | | | 4.4 | 3.8 | | | 4.8 | 2.8 | | | |
| H. Dundo (Maternid | urban | Lunda Norte | | | | | | | | | | | | | | | | | 3.4 | 3.4 | | 6.6 | 5.6 | 6.4 | | | |
| Nzage-Cambulo | rural | Lunda Norte | | | | | | | | | | | | | | | | | | | | 3.6 | 2.9 | | 3.8 | 3.8 | |
| C.M.I. Saurimo | urban | Lunda Sul | | | | | | | | | | | | | | | 1.0 | | 3.4 | 3.6 | | 4.6 | 2.8 | 4.0 | | | |
| Muconda | rural | Lunda Sul | | | | | | | | | | | | | | | | | | | | 0.8 | 2.2 | | 3.0 | 3.0 | |
| C.S. Ritondo | urban | Malange | | | | | | | | | | | | | | | | | 1.4 | 1.8 | | 1.8 | 0.8 | 1.6 | | | |
| C.M.I. Luena | urban | Moxico | | | | | | | | | | | | | | | | | 2.6 | 2.0 | | 3.2 | 3.0 | 3.0 | | | |
| Luau | rural | Moxico | | | | | | | | | | | | | | | | | | | | 2.4 | 0.6 | | 3.0 | 3.0 | |
| H. Mat.Infantil do N | urban | Namibe | | | | | | | | | | | | | | | | | 2.0 | 3.6 | | 2.1 | 3.7 | 4.0 | | | |
| Maternidade de Uig | urban | Uige | | | | | | | | | | | | | | | | | 4.9 | 1.0 | | 0.6 | 1.0 | | | | |
| Negage | rural | Uige | | | | | | | | | | | | | | | | | | | | 1.8 | 1.0 | 0.6 | | 0.6 | |
| C.S 1º de Maio Zair | urban | Zaire | | | | | | | | | | | | | | | | | 2.2 | 2.1 | | 3.6 | 1.6 | 1.4 | | | |
| H.P Zaire | urban | Zaire | | 9.0 | | | | | | | | | | | | | | | | | | 2.3 | 0.8 | 0.6 | | | |